\documentclass[a4paper,11pt]{article}
\usepackage{pos}

\usepackage{graphicx}
\usepackage{amssymb}
\usepackage{amsmath,esint}
\usepackage{lineno}
\usepackage{subfig}
\usepackage{enumitem}
\usepackage{setspace}

\usepackage{caption}

\title{A novel microstructure-based model to explain the IceCube ice anisotropy}

\ShortTitle{A novel microstructure based model to explain the IceCube ice anisotropy}

\author{The IceCube Collaboration \\{\normalsize \normalfont(a complete list of authors can be found at the end of the proceedings)}}

\emailAdd{martin.rongen@icecube.wisc.edu}
\emailAdd{dima@icecube.wisc.edu}

\abstract{
The IceCube Neutrino Observatory instruments about 1 km$^3$ of deep, glacial ice at the geographic South Pole using 5160 photomultipliers to detect Cherenkov light of charged relativistic particles. Most of IceCube's science goals rely heavily on an ever more precise understanding of the optical properties of the instrumented ice. A curious light propagation effect observed by the experiment is an anisotropic attenuation, which is aligned with the local flow of the ice. Having recently identified curved photon trajectories resulting from asymmetric light diffusion in the birefringent polycrystalline microstructure of the ice as the most likely underlying cause of this effect, work is now ongoing to optimize the model parameters (effectively deducing the average crystal size and shape in the detector). We present the parametrization of the birefringence effect in our photon propagation simulation, the fitting procedures and results as well as the impact of the new ice model on data-MC agreement. 

\vspace{4mm}
{\bfseries Corresponding authors:}
Martin Rongen$^{1*}$, Dmitry Chirkin$^{2}$\\
{$^{1}$ \itshape Institute of Physics, University of Mainz, D-55099 Mainz, Germany}\\
{$^{2}$ \itshape Dept. of Physics and WIPAC, University of Wisconsin, Madison, WI 53706, USA}\\[4mm]
$^*$ Presenter
}

\FullConference{37$^{\rm{th}}$ International Cosmic Ray Conference (ICRC 2021)\\
		July 12th -- 23rd, 2021\\
		Online -- Berlin, Germany}

\begin{document}
\maketitle

\section{Introduction}

The IceCube Neutrino Observatory is a cubic-kilometer neutrino  detector instrumenting depths between 1450\,m and 2450\,m in the ice at the geographic South Pole \cite{detector:paper}. Neutrino event reconstruction relies on the optical detection of Cherenkov radiation emitted by charged secondary particles produced in neutrino interactions in the surrounding ice or the nearby bedrock. The optical properties of ice surrounding the photo sensors are described with a table of absorption and effective scattering coefficients approximating average ice properties in 10\,m-thick ice layers. These properties were determined with dedicated calibration measurements as described in \cite{Aartsen2013}.

\begin{figure}[h]
    \centering
    \includegraphics[width=0.70\textwidth]{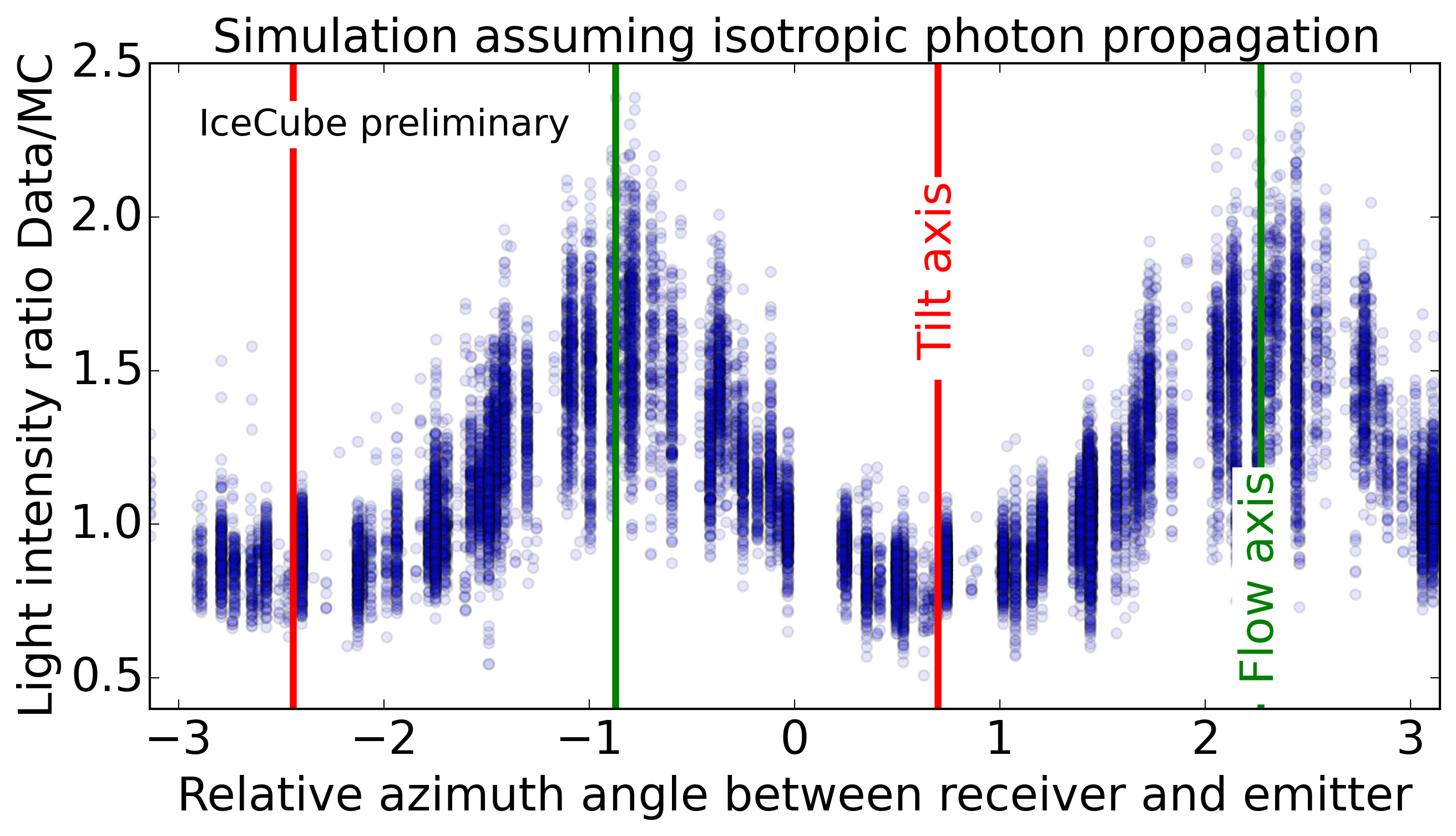}
    \caption{Optical ice anisotropy seen as azimuth dependent intensity excess in flasher data.}
    \label{fig:ratio}
\end{figure}

In the calibration runs, all functional ($>98\%$) sensors (digital optical modules, or DOMs) of the detector were operated in ``flasher'' mode (one at a time) to emit light from  built-in LEDs  in approximately azimuthally-symmetric patterns. The emitted light was then observed by the DOMs on the surrounding strings.
As previously reported \cite{ICRC13_anisotropy,TC:logger}, ice at the South Pole exhibits a strong anisotropy in light propagation at macroscopic scales. The ice flows (moves) in the direction grid NW at a rate of about 10 m/year. Layers with roughly constant scattering and absorption change in depth by as much as 60\,m as one moves across the $\sim 1$\,km detector, mainly in the SW gradient direction grid of the bedrock (this is called "tilt"). The anisotropy effect appears to align very well with the tilt gradient and ice flow directions. Measured at $\sim$125\,m from an isotropic emitter (averaging over many flashers), about twice as much light reaches DOMs along the flow axis than on the orthogonal tilt axis (see Figure \ref{fig:ratio}). At the same time the arrival time distributions are nearly unchanged compared to a simulation expectation without anisotropy.

\begin{figure}[h]
    \centering
    \includegraphics[width=0.48\textwidth]{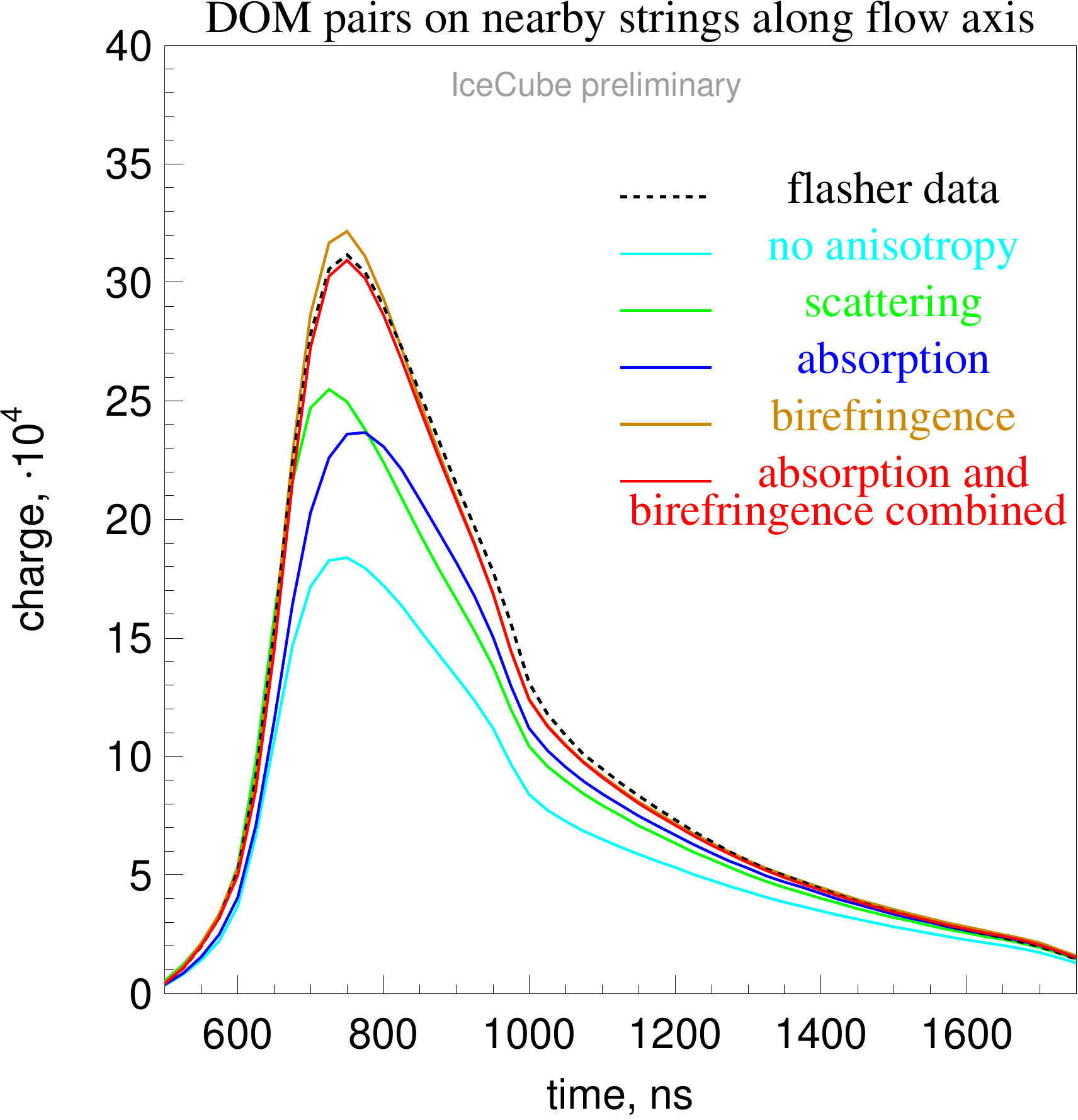}\hfill
    \includegraphics[width=0.48\textwidth]{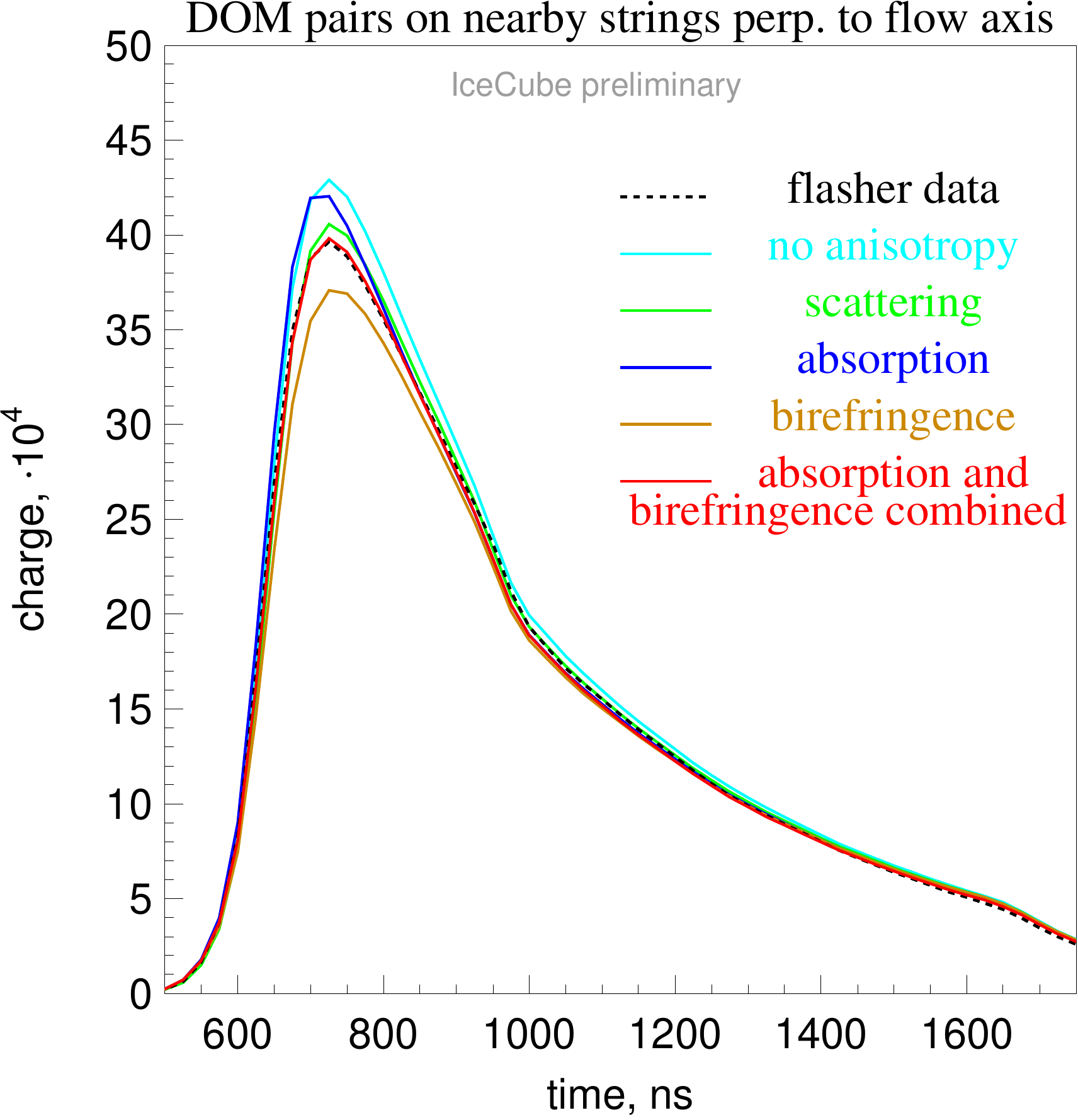}
    \caption{Comparison of fit quality achieved with different models of anisotropy as described in this report. Shown are combined photon arrival time distributions for all nearest-pair emitters and receivers, roughly aligned along and perpendicular to the ice flow.}
    \label{fig:lightcurves}
\end{figure}

Several parametrizations modifying the scattering function, absorption and / or scattering coefficients as a function of propagation angle have been explored in the past with some success \cite{ICRC13_anisotropy,Rongen:PHD} (see Figure \ref{fig:lightcurves}). However, none of them were able to fit intensity and timing distributions simultaneously. Departing from the paradigm that the scattering of light in ice is only caused by particulate impurities, light deflection resulting from asymmetric diffusion in birefringent ice polycrystals with preferential c-axes distributions was proposed in 2019 \cite{ICRC19:anisotropy}. 

Building on this work, we present the first IceCube ice model fitted to LED data, called SpiceBFR, which incorporates this effect.

\section{Simulating diffusion patterns}

Simulating each crystal/grain boundary crossing and the resulting refractions and reflections over the full, kilometer scale required during photon propagation is computationally unfeasible using the software introduced in \cite{ICRC19:anisotropy}. Thus ensembles of photon positions and directions for a given initial photon direction and crystal realization in otherwise optically perfect ice, called diffusion patterns, are simulated for a fixed number of grain boundary crossings. This number has, in this work, been chosen to be 1000 which equates to roughly 1\,m of ice. These diffusion patterns are then parametrized in terms of their diffusion, defection and displacement as explained in the following section and applied in photon propagation as explained in section \ref{section:implementation}. 

Considering Snell's law, both the encountered refractive indices as well as the boundary surface orientations dictate the resulting diffusion patterns. The refractive index experienced by the extraordinary rays depends on the opening angle between the wave vector and the crystal axis of the traversed grain. Thus the distribution of c-axes found in many crystals, also referred to as fabric in glaciology, needs to be modeled and sampled from for each simulated grain. This is realized as described in \cite{TC:caxes}, with the characteristics parameters $\ln(S_1/S_2)$ and $\ln(S_2/S_3)$ as also used and measured during ice core analysis.

As the average grain shape deviates from a sphere, the encountered distribution of face orientations depends on the photon direction. Assuming the face orientation of a solid, tessellated into elongated polyhedra, to be described by the surface orientation density of an ellipsoid describing the average grain shape, one can sample the distribution from analytic functions as described in \cite[p.~169]{Rongen:PHD}.
Early investigations concluded that spheroids with their mayor axis being aligned to the flow direction are strongly preferred over arbitrary ellipsoids. Thus we here restrict ourselves to this case, simplifying both the simulation as well as the parametrization, as deflection and diffusion now simply depend on the opening angle between the photon direction and the ice flow axis.

While elongation and fabric are intrinsically linked, no quantitative relationship between the two is know to us, thus they are treated as independent free parameters. Equally no correlation between the change in c-axis and the grain boundary inclination of neighboring grains is assumed in the simulation.

\section{Parametrizing diffusion patterns}
\label{section:parametrization}

\begin{figure}
\centering
\begin{minipage}{.48\textwidth}
    \centering
    \includegraphics[width=\linewidth]{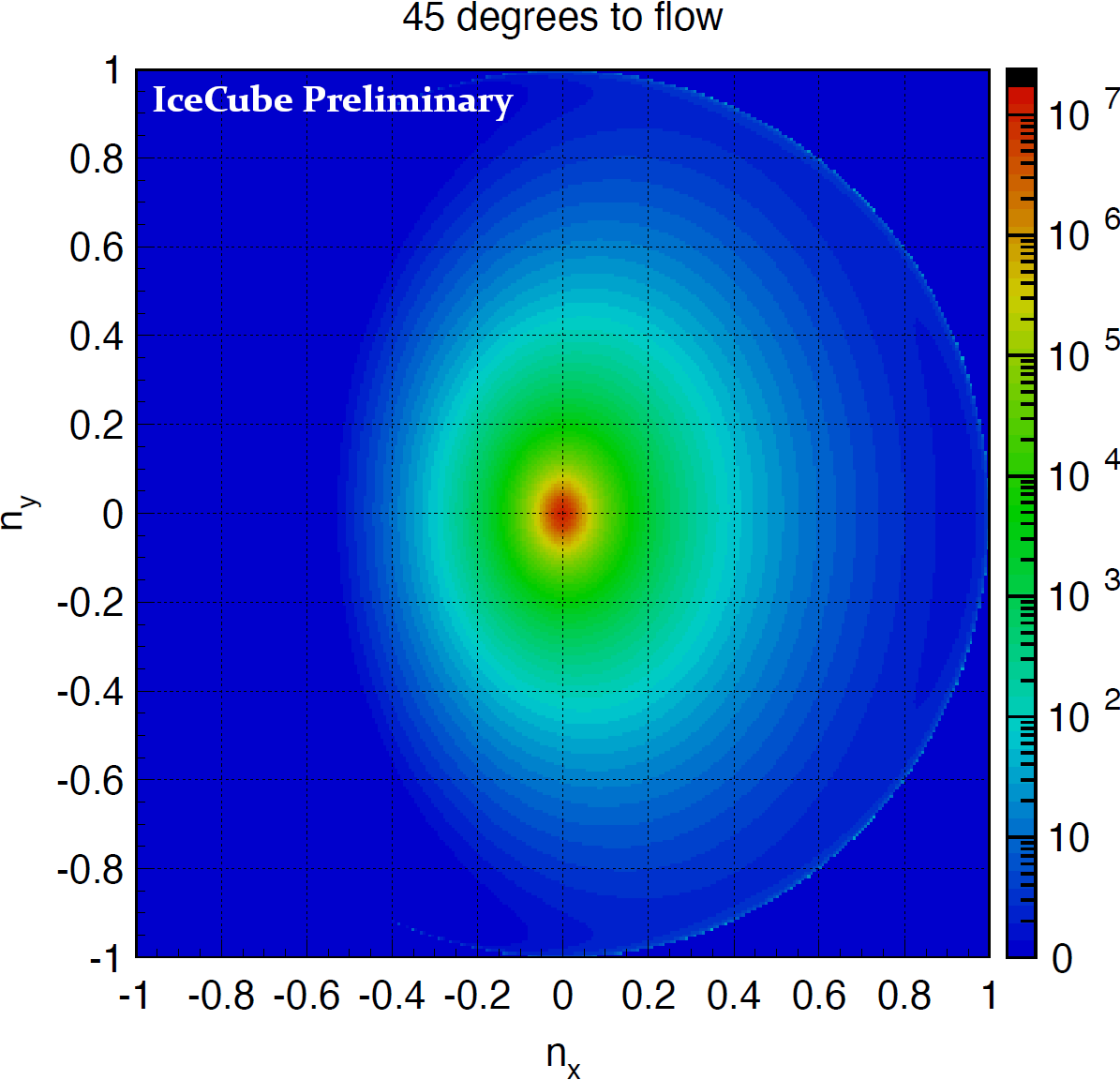}
    \caption{Example of a diffusion pattern  after photon propagation through 1000 crystals. Initial direction perpendicular to the picture. The subtle effect of photon scattering towards the ice flow (with components (1,0,1)/$\sqrt{2}$ in this example) can be seen.}
    \label{fig:pattern}
\end{minipage} \hfill %
\begin{minipage}{.48\textwidth}
    \centering
    \includegraphics[width=\linewidth]{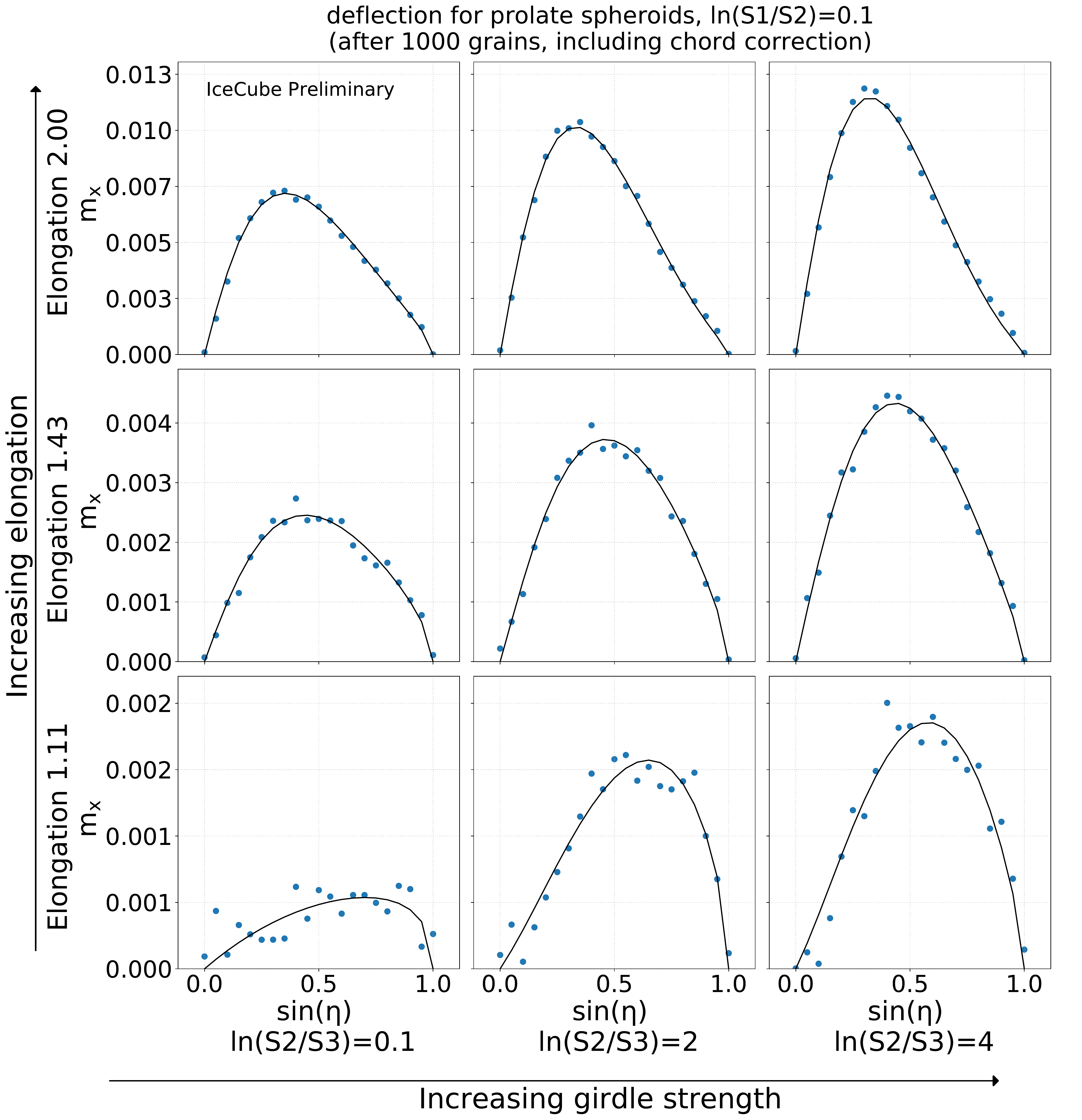}    \caption{Deflection as a function of opening angle to the flow for a number of crystal configurations. The black curves were fitted through the blue, simulated points.}
    \label{fig:deflections}
\end{minipage}
\end{figure}

Diffusion patterns have been simulated for a wide range of spheroid elongations and fabric parameters. As evident from the example in Figure \ref{fig:pattern}, these diffusion patterns have a strong central core with a wide tail dominated by mainly single large angle reflections. As such, the tail scales linearly with number of crystal crossings. We found that the precise simulation of the tail is unimportant; therefore the distribution is modeled as a 2d-Gaussian on a sphere, lending itself to usual scaling (with distance) relationships for mean displacement and width. The distributions are very slightly skewed towards the flow axis
. A number of more complicated fit functions were also tried with good success in precisely describing the underlying distribution (in fact this is what is shown in Figure \ref{fig:pattern} to illustrate all features of the distribution without statistical fluctuations). These were however abandoned, as no simple distance scaling could be established.

The three parameters of the diffusion pattern modeled with the 2d-Gaussian on a sphere are the two widths (in the directions towards the flow, $\sigma_x$, and perpendicular to it, $\sigma_y$), and a single mean deflection towards the flow, $m_x$. Mean deflection in the perpendicular direction was zero for all cases that we chose to include into the final model (i.e., single-axis ellipsoids for particle shape and selected crystal fabric configurations). Because we mainly simulate small deflections (ignoring the long tails), we simulated the 2d Gaussian in Cartesian coordinates, and then projected that to the sphere with an inverse stereographic projection. The three quantities were fitted to the following functions of angle $\eta$ of the initial photon direction with respect to the ice flow, for simulations with a fixed number of 1000 crystal crossings:
\begin{equation}
    \sigma_{x,y} = A \cdot \exp(-B \cdot (\arctan(D \sin \eta))^C) \quad (\mbox{diffusion})
\end{equation}
\begin{equation}
    m_x = A \cdot \arctan(D\cdot \sin \eta \cos \eta) \cdot \exp(-B \sin \eta + C \cos \eta) \quad (\mbox{deflection}).
\end{equation}

These functions were found by trial and error and describe all considered crystal realizations with only 12 parameters. Figure \ref{fig:deflections} shows the mean deflection for nine crystal configurations. Note that increasing elongation has a stronger effect compared to strengthening fabric. 

\section{Applying diffusion patterns in photon propagation}
\label{section:implementation}

During normal photon propagation directions are only updated upon scattering. To minimize the additional computational burden, the new birefringence anisotropy is discretized and also evaluated only at the scattering sites. This requires scaling the diffusion, deflection and displacement derived from simulation through 1000 grains to the number of traversed grains between two scattering sites. This introduces a new model parameter, the average grain size and also requires taking into account the different average crystal chord lengths as a function of propagation direction (as described in \cite{Rongen:PHD}), further increasing the importance of elongation over fabric.
As would be expected from a diffusion process, and was confirmed in simulation, the deflection scales linearly and the diffusion as the square root of traversed grains $n$ ($\sigma_{x,y}\propto \sqrt{n}$ \& $m_x\propto n$).
To decouple the fitting of anisotropy properties from the overall ice diffusion / scattering strength as much as possible, the effective scattering Mie coefficient was reduced by the amount resulting from the birefringence induced light diffusion assuming on average isotropic photon directions.

Although missing from early fits, updating not only a photon's direction with deflection due to birefringence, but also the photon coordinates (as it shifts transversely with respect to straight-path expectation), improves the quality of description of data in the final fit. Due to the simple physics of cumulative photon deflections, the effect can be simulated at a small additional computational cost and with no additional parameters. Assuming WLOG that all birefringence deflections happen at constant distance interval $\Delta l$ and that these can be sampled from the same distribution (which depends on the initial photon direction), as the individual and even final calculated deflections are very small, we can express the new photon direction and coordinates after $N$ deflections as:
\begin{equation}
    \vec{n}=\vec{n}_0+\sum_{i=1}^N{\Delta\vec{n}_i}, \quad \vec{r}=\sum_{i=1}^N{\vec{n}_i\cdot \Delta l}=\Delta l\cdot N\cdot\vec{n_0}\cdot+\Delta l\cdot\sum_{i=1}^N{\sum_{j=1}^i{\Delta \vec{n}_j}}
\end{equation}

Therefore the second term in each of the two expressions above describes a cumulative direction change $\delta\vec{n}$ and relative coordinate update $\delta\vec{r}$ respectively (we note that the total distance traveled is $L=\Delta l\cdot N$). We can now calculate, that in the limit or large $N$,
\begin{equation}
    <\delta \vec{r}>=<\delta \vec{n}>\frac{L}{2}, \quad
    <\Delta(\delta\vec{r}-\delta\vec{n}\frac{L}{2})^2>=<\Delta(\delta\vec{n})^2>\cdot \frac{L^2}{12}, \quad
    <\Delta(\delta\vec{r}-\delta\vec{n}\frac{L}{2})\cdot \Delta(\delta\vec{n})>=0.
\end{equation}

These equations indicate that the coordinate update $\delta\vec{r}$ can be sampled from a distribution with a mean given by the first equation 
and variance given by the second equation. Because there is no correlation between the residual in the variance and the deflection vector, as shown by the third equation, the variance can be sampled using the already tabulated birefringence parameters independently from sampling the variance of the deflection vector.

\section{Fitting to flasher data} \label{section:fitting}

The model described above requires four parameters to specify a birefringence anisotropy realization, i.e. crystal size \& elongation and two fabric parameters). Additionally allowing for a correction to the previously established Mie absorption and scattering coefficients adds two more parameters. As minimizing all six parameters for all 100 depth layers in the ice model is not computationally feasible, we need to identify some which are either depth independent or have a small effect on the data-MC agreement.

\begin{figure}[h]
    \centering
    \includegraphics[width=0.975\textwidth]{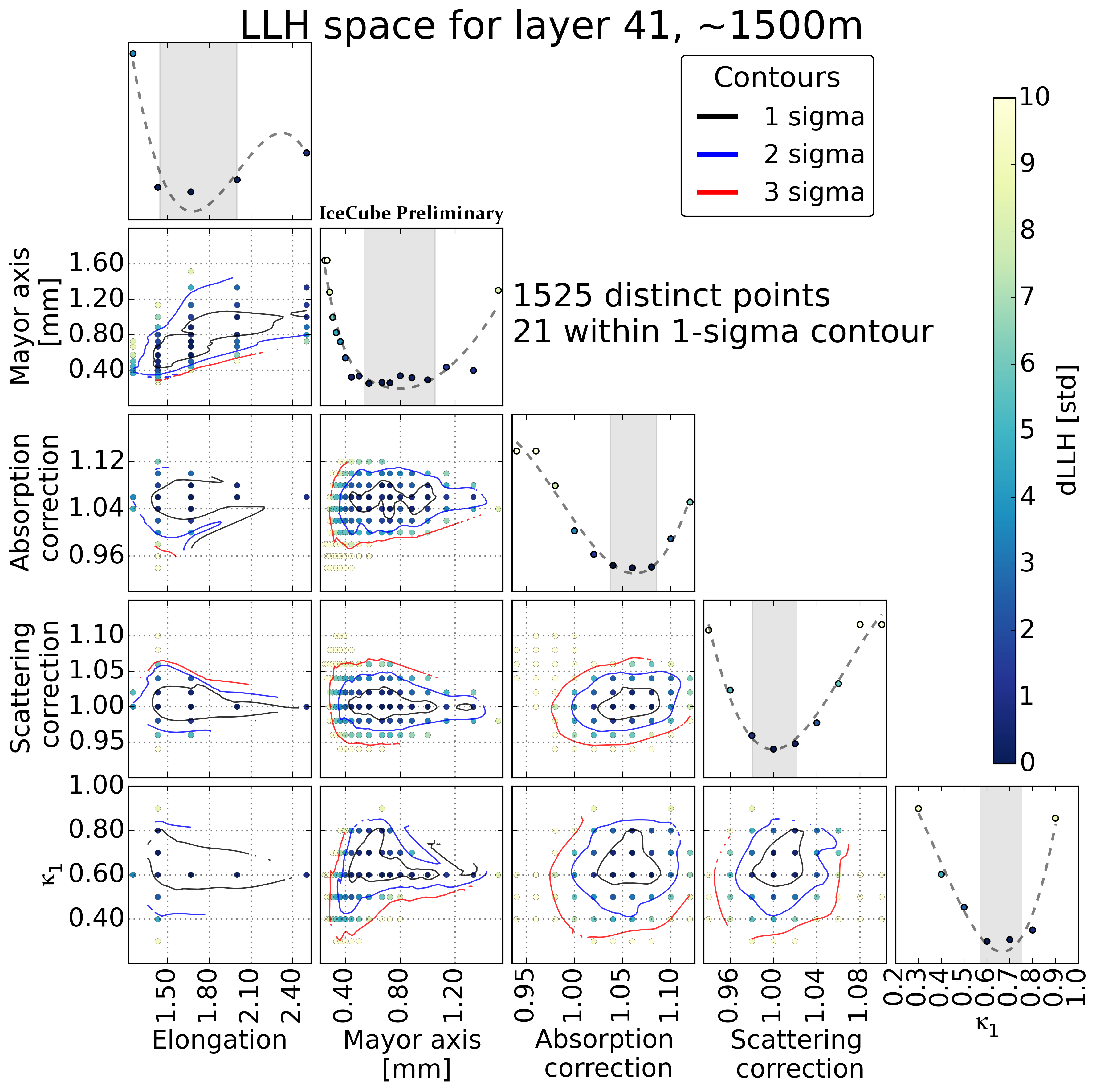}
    \caption{Example LLH space for one ice layer and a subset of parameters. Each panel shows a marginalized 2D space, each point being a simulated ice realization, color coded by its LLH distance from the best fit.}
    \label{fig:fit}
\end{figure}

The required pre-fits, as well as the final depth evaluation, were performed following the method described in \cite{Aartsen2013} by minimizing the summed LLH\footnote{The minus log-likelihood, denoted here as LLH, is akin to the saturated Poisson likelihood, and can similarly be used as a measure of the goodness-of-fit.} comparing the {\it single-LED} data set (where all 12 LEDs on all in-ice DOMs were flashed one at a time) with the full photon propagation simulation of these events taking into account precisely known DOM orientations as measured in \cite{Dima2021}. 
Fits for individual layers were carried out by only including LEDs situated within the considered ice layer into the LLH summation. This method offers a reduced depth resolution compared to \cite{Aartsen2013}, but reduces computation time while making use of the full data. An example LLH space at a depth of $\sim$1500\,m is shown in Figure \ref{fig:fit}.  During the pre-fits the following behavior was noted:
Given a girdle fabric ($\ln(S_1/S_2) >> \ln(S_2/S_3)$), the actual fabric strength has a small effect and can not be distinguished by the data. Accordingly the fabric has been fixed to values as measured in the deepest sections of the South Pole Ice Core, SPC14, \cite{SPC14:caxis} ($\ln(S_1/S_2)=0.1 \,\&\, \ln(S_2/S_3)=4$).
The fit is largely degenerate in crystal elongation and size, with small, near spherical crystals yielding similar results to larger, more elongated realizations. Thus, the elongation was fixed to 1.4, which is a good fit at all layers and similar to the values as measured in the deepest parts of SPC14\cite{Alley2021}.

Fitting the remaining parameters, crystal size and absorption \& scattering correction for all layers, yields a significant improvement as seen for example in the average light curves in Figure \ref{fig:lightcurves} (birefringence only line). Still the best-fit does not perfectly match the data and more worryingly the required crystal sizes are on the order of 0.1\,mm and as such far smaller than expected from glaciological literature\cite{Alley2021}. 
After thoroughly checking both the assumptions and implementation of the birefringence model, we decided to reintroduce scattering as well as absorption anisotropy, both following the formalism as described in \cite{ICRC13_anisotropy}, into the fit. As would be expected from the timing behavior, the fit does not make use the scattering anisotropy, but surprisingly the absorption anistropy is mixed into the birefringence model with a significant non-zero contribution (nearly depth independent with an average of $\kappa_1=0.6 \,\&\, \kappa_2 = \kappa_3 = -0.3)$. This means a departure from a first-principle model, but was adopted for its improvement in data-MC agreement. After including the absorption anisotropy, crystal size and absorption \& scattering correction were again fitted for all layers. 

\section{Resulting ice model}
\label{section:results}

\begin{figure}
\centering
\begin{minipage}{.48\textwidth}
    \centering
    \includegraphics[width=\linewidth]{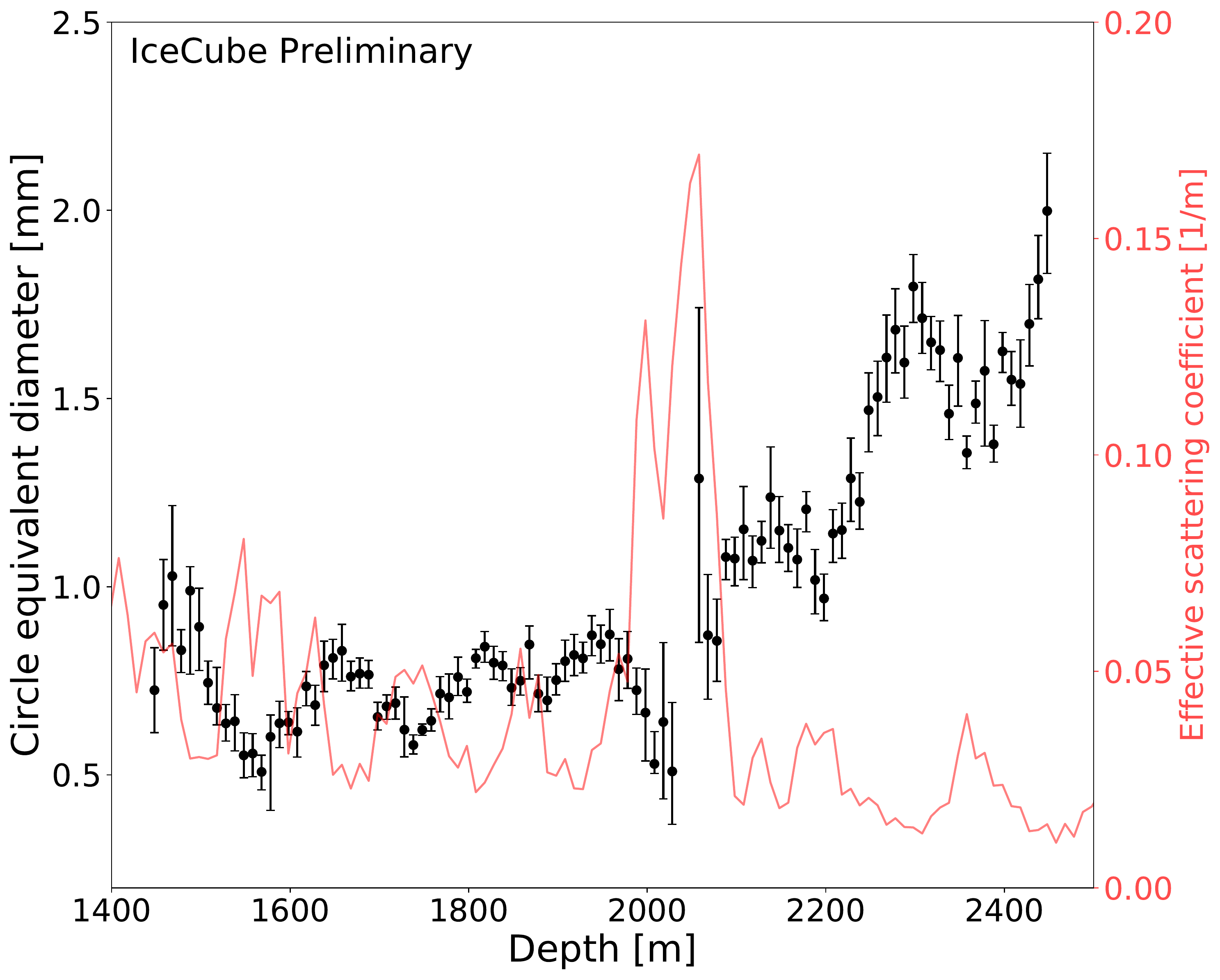} 
    \caption{Best fit crystal sizes. Error bars denote the statistical uncertainty only.}
    \label{fig:size}
\end{minipage} \hfill %
\begin{minipage}{.48\textwidth}
    \centering
    \includegraphics[width=\linewidth]{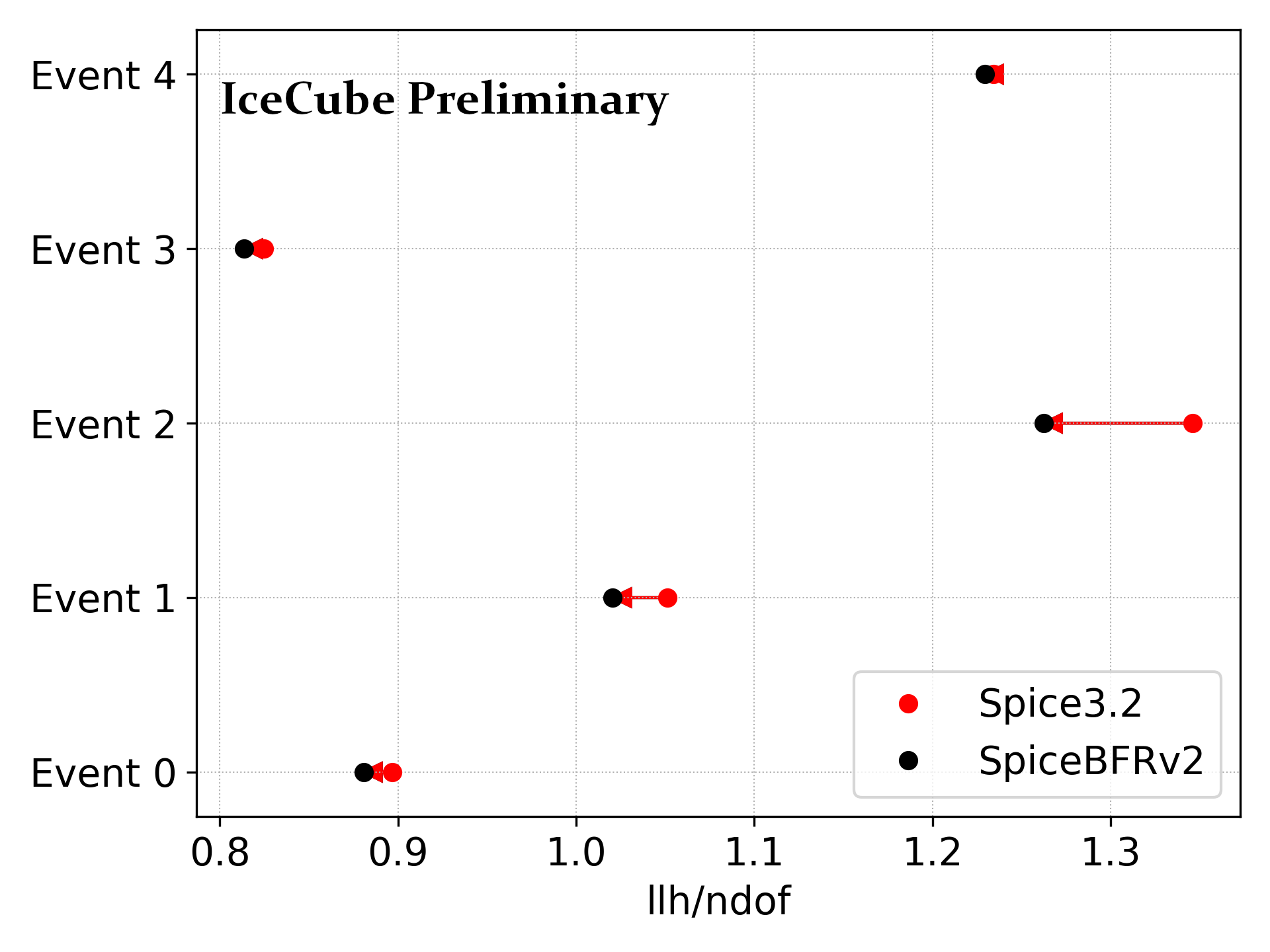} 
    \caption{Direct fit results for five example high energy events. Reconstructing with SpiceBFR leads to an improved data description in all cases.}
    \label{fig:DF}
\end{minipage}
\end{figure}

Figure \ref{fig:size} depicts the best fit stratigraphy of grain sizes. The overall grain size of $\sim$1\,mm as well as the increase in older and cleaner ice are as generally expected and measured in glaciology \cite{EPICA2004, Alley2021}. 

As seen in Figure \ref{fig:lightcurves} the new model significantly improves in matching the flasher data light curves both in terms of timing and total intensity with regards to older models and for the first time achieves an excellent data-MC agreement.
Wide spread application in physics analyses requires large scale simulations and is still in preparation. Still, first tests employing the ice model in direct fit reconstructions \cite{Chirkin:2013avz} of HESE events as exemplified in Figure \ref{fig:DF} confirm that the improved data-MC agreement seen in flasher data also translates to better reconstructions.

\section{Summary and Outlook}
\label{section:summary}

A model combining anisotropic absorption with light deflection resulting from propagation through the birefringent ice polycrystal significantly improves on previous ice models. The model yields a near perfect data-MC agreement for flasher data in timing and intensity variables and will improve event reconstructions while also reducing systematic biases.  
In the fitting process the average crystal size in the detector and its correlation to impurity concentrations, as relevant in glaciology, is deduced. While the birefringence model has been deduced from first principle, the absorption contribution is so far unmotivated. Understanding its origin or extending the birefringence to absorb the full effect will be the focus of future studies, in particular in the IceCube Upgrade.

\bibliographystyle{ICRC}
\bibliography{sample}

\clearpage
\section*{Full Author List: IceCube Collaboration}




\scriptsize
\noindent
R. Abbasi$^{17}$,
M. Ackermann$^{59}$,
J. Adams$^{18}$,
J. A. Aguilar$^{12}$,
M. Ahlers$^{22}$,
M. Ahrens$^{50}$,
C. Alispach$^{28}$,
A. A. Alves Jr.$^{31}$,
N. M. Amin$^{42}$,
R. An$^{14}$
% ...
\\

\noindent
% ...
$^{12}$ Universit{\'e} Libre de Bruxelles, Science Faculty CP230, B-1050 Brussels, Belgium \\
% ...
$^{14}$ Department of Physics and Laboratory for Particle Physics and Cosmology, Harvard University, Cambridge, MA 02138, USA \\
% ...
$^{17}$ Department of Physics, Loyola University Chicago, Chicago, IL 60660, USA \\
$^{18}$ Dept. of Physics and Astronomy, University of Canterbury, Private Bag 4800, Christchurch, New Zealand \\
% ...
$^{22}$ Niels Bohr Institute, University of Copenhagen, DK-2100 Copenhagen, Denmark \\
% ...
$^{28}$ D{\'e}partement de physique nucl{\'e}aire et corpusculaire, Universit{\'e} de Gen{\`e}ve, CH-1211 Gen{\`e}ve, Switzerland \\
% ...
$^{31}$ Karlsruhe Institute of Technology, Institute for Astroparticle Physics, D-76021 Karlsruhe, Germany  \\
% ...
$^{42}$ Bartol Research Institute and Dept. of Physics and Astronomy, University of Delaware, Newark, DE 19716, USA \\
% ...
$^{50}$ Oskar Klein Centre and Dept. of Physics, Stockholm University, SE-10691 Stockholm, Sweden \\
% ...
$^{59}$ DESY, D-15738 Zeuthen, Germany \\
% ...

\subsection*{Acknowledgements}

\noindent
USA {\textendash} U.S. National Science Foundation-Office of Polar Programs,
U.S. National Science Foundation-Physics Division,
U.S. National Science Foundation-EPSCoR,
Wisconsin Alumni Research Foundation,
Center for High Throughput Computing (CHTC) at the University of Wisconsin{\textendash}Madison,
Open Science Grid (OSG),
Extreme Science and Engineering Discovery Environment (XSEDE),
Frontera computing project at the Texas Advanced Computing Center,
U.S. Department of Energy-National Energy Research Scientific Computing Center,
Particle astrophysics research computing center at the University of Maryland,
Institute for Cyber-Enabled Research at Michigan State University,
and Astroparticle physics computational facility at Marquette University;
Belgium {\textendash} Funds for Scientific Research (FRS-FNRS and FWO),
FWO Odysseus and Big Science programmes,
and Belgian Federal Science Policy Office (Belspo);
Germany {\textendash} Bundesministerium f{\"u}r Bildung und Forschung (BMBF),
Deutsche Forschungsgemeinschaft (DFG),
Helmholtz Alliance for Astroparticle Physics (HAP),
Initiative and Networking Fund of the Helmholtz Association,
Deutsches Elektronen Synchrotron (DESY),
and High Performance Computing cluster of the RWTH Aachen;
Sweden {\textendash} Swedish Research Council,
Swedish Polar Research Secretariat,
Swedish National Infrastructure for Computing (SNIC),
and Knut and Alice Wallenberg Foundation;
Australia {\textendash} Australian Research Council;
Canada {\textendash} Natural Sciences and Engineering Research Council of Canada,
Calcul Qu{\'e}bec, Compute Ontario, Canada Foundation for Innovation, WestGrid, and Compute Canada;
Denmark {\textendash} Villum Fonden and Carlsberg Foundation;
New Zealand {\textendash} Marsden Fund;
Japan {\textendash} Japan Society for Promotion of Science (JSPS)
and Institute for Global Prominent Research (IGPR) of Chiba University;
Korea {\textendash} National Research Foundation of Korea (NRF);
Switzerland {\textendash} Swiss National Science Foundation (SNSF);
United Kingdom {\textendash} Department of Physics, University of Oxford.

\end{document}